# Human-in-the-loop: The future of Machine Learning in Automated Electron Microscopy

Sergei V. Kalinin,[1,1] Yongtao Liu, [2]Arpan Biswas,[2] Gerd Duscher,[1] Utkarsh Pratiush,[1] Kevin Roccapriore,[2] Maxim Ziatdinov,[2] and Rama Vasudevan[2]

[1] Department of Materials Science and Engineering, University of Tennessee, Knoxville, TN 37996, USA

[2] Center for Nanophase Materials Sciences, Oak Ridge National Laboratory, Oak Ridge, TN 37831, USA

Machine learning methods are progressively gaining acceptance in the electron microscopy community for de-noising, semantic segmentation, and dimensionality reduction of data post-acquisition. The introduction of the APIs by major instrument manufacturers now allows the deployment of ML workflows in microscopes, not only for data analytics but also for real-time decision-making and feedback for microscope operation. However, the number of use cases for real-time ML remains remarkably small. Here, we discuss some considerations in designing ML-based active experiments and pose that the likely strategy for the next several years will be human-in-the-loop automated experiments (hAE). In this paradigm, the ML learning agent directly controls beam position and image and spectroscopy acquisition functions, and human operator monitors experiment progression in real- and feature space of the system and tunes the policies of the ML agent to steer the experiment towards specific objectives.

[1] sergei2@utk.edu

The motivation for this opinion piece stems from the July 2023 Microscopy Society Meeting in Minneapolis. One of the hallmarks of the meeting was the large number of presentations on machine learning (ML) in microscopy, ranging from denoising, unsupervised data analysis via variational autoencoders, and supervised learning applications for semantic segmentations and feature identification. Remarkably, by now most manufacturers offer or have plans to offer Python application programming interfaces (APIs), allowing the deployment of the codes on operational microscopes. From this perspective, the technical barriers for the broad implementation of automated microscopy in which ML algorithms analyze the data streaming from instrument detectors and make decisions based on this data are lower than ever. However, the rate of adoption of ML solutions as a part of real-time microscope operations is extremely low. Here we offer some considerations on what the limiting steps in this process are.

The current probe microscopy paradigm – whether it is scanning electron microscopy, scanning transmission electron microscopy, or variants of scanning probe microscopy – has not changed since the very dawn of these techniques. Like in 80ies and 90ies, we use standard rectangular scans to acquire structural information and use the rectangular spatial grids for hyperspectral imaging. The experiment orchestration and execution – that is, selection of the specific locations for imaging and spectroscopy - is driven by a human operator, much like in Ernst Ruska's time.

Can ML – driven microscopy change this paradigm, and what is the value proposition in introducing ML workflows? For intrinsically slow techniques such as scanning tunneling microscopy, the automatization of microscope operation can liberate the operator from tedious work and improve traceability and reproducibility of the data acquisition process but will not significantly increase the amount of data from the instrument. Combined with complex logic used by human operators during decision-making that is difficult to reproduce in the ML workflow and small number of such tools, this will make the value of ML adoption relatively low.

The situation in the field of electron microscopy is completely different. Here, the sustained progress in data acquisition electronics and detectors over the last two decades now allows data acquisition in volumes and rates well above human analysis capabilities or reaction speeds. For example, 32k pixel images can be acquired - but given the limited perception field of the human eye, decision-making based on this data is complicated. Similar considerations apply to the hyperspectral EELS and 4D STEM data. Here, the high-dimensional data needs to be converted to 2D representations to make decisions, and this conversion can be accomplished in multiple different ways and typically **after** the experiment. These considerations must be further positioned in the context of the size of electron microscopy market – that now approaches 1.7$B and covers areas from semiconductor industry to biology and drug target development and virtually all areas of fundamental research from condensed matter physics to catalysis.

The current trends followed by the instrumental manufacturers are a focus on improving spatial and energy resolution of the methods (always a good target) and methods to minimize beam damage (lower voltage, dose monitoring, equal sampling). For the latter, these are the only approaches that can be proposed from a problem agnostic perspective, i.e. are expected to be equally important for any material system. This is similar to the classical hyperspectral imaging strategies – sampling on a uniform spatial grid is an optimal strategy if no other information about object of interest is available.

However, from the experimentalist perspective, the objective of an experiment is to learn something about the sample in terms of structure, properties, or physics. Higher stability, energy, and spatial resolution are key factors determining the baseline of instrument performance. However, for each defined problem there will be much more effective ways to collect data than uniform grid sampling – except that it will depend on the experiment objective. For example, in the samples containing grain boundaries, dislocations, and second phase inclusions it is the defects that represent object of interest, whereas the bulk of material away from defects has fairly low value. The relative importance of the possible defect types in turn depends on the specific research objective – e.g., understanding of corrosion behavior vs. mechanical deformation. Ultimately, the customer problem is exploring specific materials problems rather than resolution or stability.

**II. Experimental Objectives and Rewards**

Traditionally, a microscope is controlled by the human operator who dynamically plans the experiment. The planning includes sample selection and preparation prior to the microscope session, decision-making of microscope tuning, selecting regions of interest, measurement parameters, etc. during microscope operation. This decision-making is guided by the combination of prior knowledge and training, curiosity, and perceived objective of the experiment. Detailed analysis is often performed after the experiment, including the initial data analysis and its interpretation in the context of the experimental goal. It can be expected that ML-driven workflows will emulate the elements of the analysis and decision-making in real-time.

The development of automated experiment workflows requires the consideration of two key concepts – **objective** and **reward**. Generally, objective refers to the non-formalized overall goal of the experiment and is not available during or by the end of the experiment. For example, exploring the hybrid perovskites for photovoltaic applications, the global objective can be mitigation of global warming effects or establishing a green energy source. The more local objective that by now is recognized to be a critical bottleneck is the optimization of stability and manufacturability of these materials.

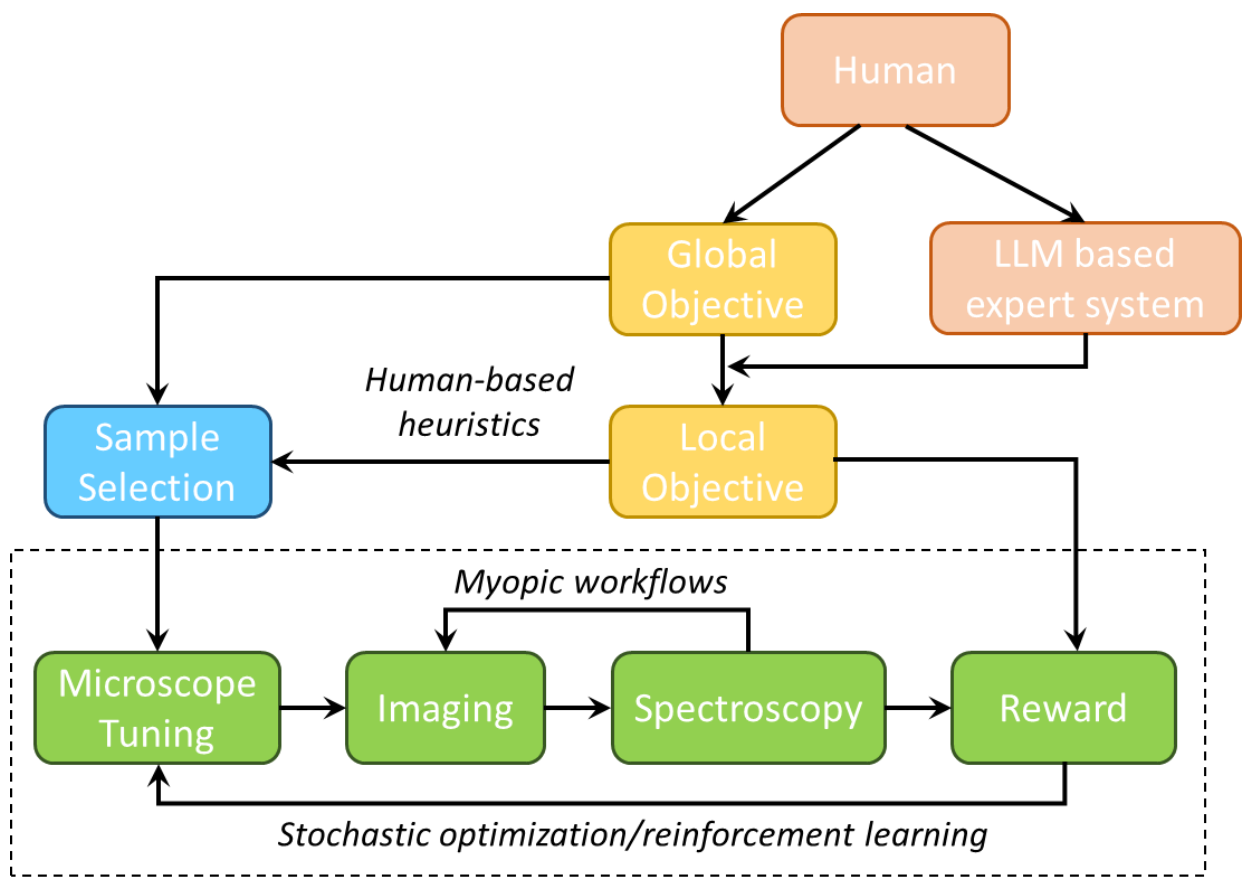

**Figure 1.** Objective and reward in electron microscopy experiment. Application of machine learning methods for workflow design generally requires specific reward to be available either at each step (myopic workflows), or after several steps/at the end of experiment. Bayesian optimization methods require reward being available at each step,

whereas reinforcement learning algorithms can be used when reward is sparse, i.e. becomes available after several steps or in the end of the experiment. LLM based expert systems can assist human operator to translate the global objective to reward function.

However, the limitation of the **objective** is that the degree to which the experiment has contributed to it often becomes obvious much later, and typically as the result of knowledge integration between multiple experiments and theory. Similarly, often experiments result in serendipitous discoveries, i.e. contribute to the objectives that were not the part of original motivation behind setting the experiment. Needless to say, these types of objectives are not suitable for the design of the ML-driven workflows.

Conversely, **reward** defines the measure of the experimental outcome that can be quantified immediately. If the reward is available at the end of the experiment or after several steps, this is the case of stochastic optimization. Here, the most well-known approach is reinforcement learning, albeit other frameworks for stochastic optimization exist.[1] If the reward is available at each experimental step, myopic optimization approaches such as Bayesian optimization can be used. The typical examples of the reward from the microscopy perspective can be resolution optimization and minimization of beam damage. From the materials discovery perspective it can be establishing structure-property relationships such as the relationship between the nanostructures and EELS spectra,[2] microstructures and polarization hysteresis loops,[3] microstructure and nonlinearity,[4] or discovery of the physical laws controlling material dynamics.[5, 6]

The relationship between the experiment objective and rewards can be extremely complicated. In fact, this is common for all active learning methods, and behaviors such as reward hacking are well recognized in the reinforcement learning community. Here, technologies like large language models such as ChatGPT can be potentially used to establish the relationship between objective and reward (for example – what should we study to understand the plasticity?). However, in general, experimental planning in the sense of defining sparse and myopic rewards given the experimental objective is likely to be a key aspect of ML incorporation in the experimental workflows.

**Table 1. Rewards**

| Reward | Definition | Example |
|---|---|---|
| Myopic | Available at each workflow step | Maximize resolution, minimize probe damage |
| Reward | Available sparsely, e.g. only at the end of an experiment | Learn structure-property relationships in materials, explore polarization switching near specific defects, discover defects responsible for low nucleation bias |
| Objective | Can be defined only in the context of more general workflows or later in time | Optimize material for specific application or address global warming |

**III. Automated experiment: static and dynamic policies**

The second key set of definitions required for defining the automated microscopy workflow is ***policy***. Essentially, a policy is a rule that determines the next command to be executed to be selected based on the currently observed state of the system. For example, the policy can be to continuously check the resolution of the instrument and initiate aberration corrector tuning if the imaging resolution drops below certain value. For active experiment, the example of policy can be initiating the spectroscopic measurements once specific object is detected.

Here, it is important to disambiguate **static** and **dynamic policies**. In a static policy setting, the policies are determined prior to the experiment and are not changed during the experiment. The policy can be either deterministic or stochastic. In a deterministic policy, the unique actions are defined for a given observed system state. For example, we can use image analysis method such as peak finder or supervised machine learning methods to discover *a priori* known objects of interest and perform the spectroscopic EELS measurements on them. In stochastic policies, the action is drawn from a known list of actions with a certain probability, and the policy is learned by finding the parameters that define these distributions.

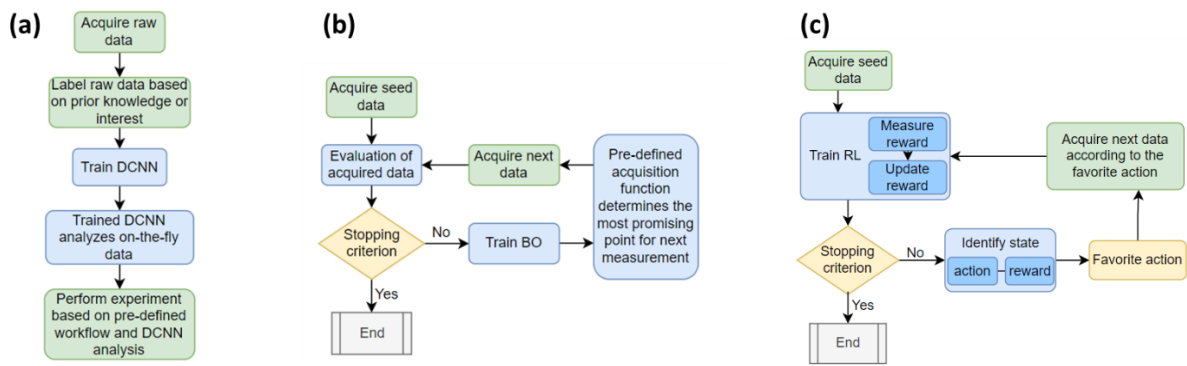

**Figure 2.** (a) Static policies in the automated experiment. (b) Myopic (single step) policies in Bayesian optimization. (c) Policies in the reinforcement learning experiment.

An example of an experiment with the static policies is the recently reported electron beam manipulation in STEM or scanning probe microscopy studies of structure-property relationships on **a-priori known** elements of interest.[7-9] In these cases, the object of interest (atoms or defect of certain type, domain walls) and actions to be performed (beam dwell for atomic assembly, spectroscopic measurements) are defined before the experiment and are not changed during the experiment.

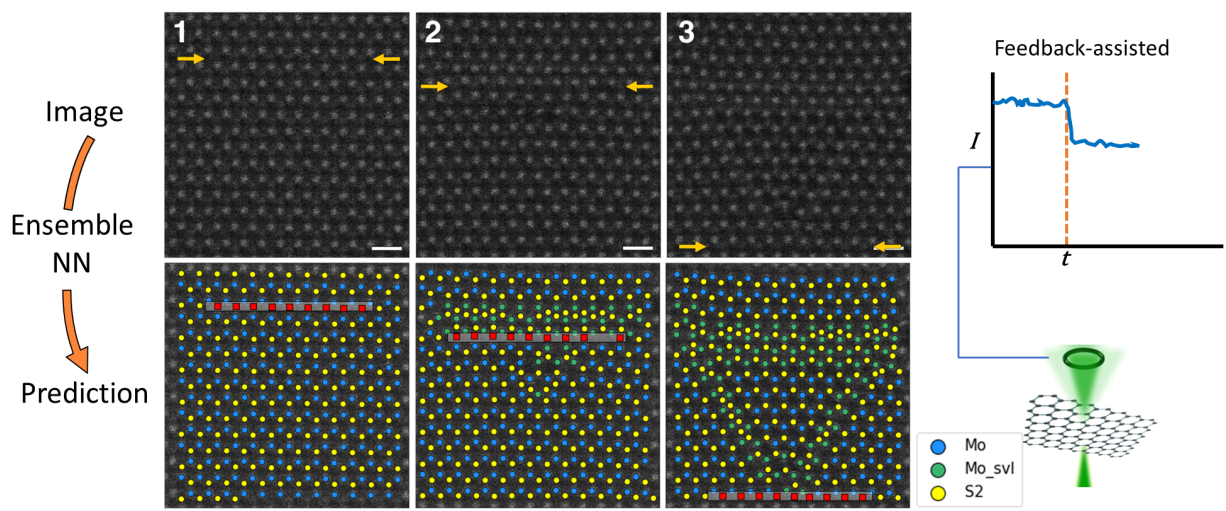

**Figure 3**. Site-specific atomic targeting as an example of fixed-policy ML driven experiment. Top row: high-angle annular dark field (HAADF)-STEM images of monolayer $MoS_2$, where the bottom row shows the classified atomic coordinate output predicted by the ensemble network; blue is Mo, yellow is $S_2$, and green are single (sulfur) vacancy line (SVL) defects. Red squares indicate the selected atomic species to be targeted by the beam, in this case, row segments of $S_2$ with intent of generating SVL defects at the desired location. Atom dwell times are modulated by the feedback of the HAADF signal, which provides selectivity to remove only one of the two S atoms in monolayer $MoS_2$ therefore allowing creation of specific quantum defects at deterministic locations. All scale bars 5 Å.

In the dynamic policy setting, the ML agent aims to learn the policies during the experiment (or during pre-experiment training in a virtual environment) to maximize the expected reward. Here, in a myopic setting the reward is available and policies are updated at each step. The example is Bayesian Optimization AE, in which policies are represented by the acquisition function that balances exploration and exploitation. Comparatively, in reinforcement learning the reward becomes available after several steps, and the goal of the algorithm is to discover policies that allow to maximize the cumulative reward.

**IV. Myopic and non-myopic workflows**

Finally, important for the discussion of the workflows is the concept of myopic and non-myopic workflows. Myopic workflows generally refer to the scenarios where the action is selected based only on the current state of the system and the reward is available after each experimental step. A typical example of myopic workflows is Bayesian optimization (BO), commonly used in ML and optimization problems where the objective function is computationally expensive to evaluate or has no analytical form. The approach involves building a probabilistic model of the objective function based on the available data and using it to select the next point to evaluate the function. This probabilistic model is updated as new data is collected, and the process is repeated until the optimal value is found. Bayesian Optimization iteratively selects candidate solutions, evaluates them, and updates the objective function based on the new information gained from each evaluation. Typical application of BO in electron microscopy will be corrector tuning or optimization of imaging conditions.

An example of machine learning methods for non-myopic workflows is reinforcement learning. Here, the aim is to learn either an explicit policy π or an implicit policy that selects which action to take at every state. The algorithm learns the policy being informed by sparse (i.e. available after several experimental steps of in the end of experiment) rewards.

The non-myopic workflows further introduce the concept of *value*. The value of each action in a workflow can be established from the reward upon its completion. In other words, the **value** is the **expected reward** from that state. For example, we aim to collect high resolution data on the DNA molecules on the surface, and we need to tune the microscope and collect the images. In this case, our reward is the number of high-quality DNA images acquired, i.e., possessing both the need to optimize imaging conditions, but also to scan a large number of molecules. However, maximizing this reward requires two steps – instrument optimization and collection of high-resolution data. Hence, the ML algorithm needs to balance what time to spend pursuing both objectives. This defines the value of the step – how much it contributes to reward.

The value assignment becomes nontrivial very rapidly with the complexity of the workflow – for multi-step workflows, the number of parameters to optimize can be very large. However, the relationship between policies and values can be established mathematically in reinforcement learning framework. In value-based methods, the policy is implicit and defined through a value function that assigns a value to each state or state-action pair. The agent selects actions that lead to higher value states, in a method such as 'Q learning.' The value of a state is taken as the average return of all trajectories from that state, assuming the same policy is pursued until termination.

In human driven workflows, value is often intuitively understood, and operators use experience in practice to make a determination about which steps are worth optimizing for what lengths of time. As an example, in the atomic force microscopy (AFM) workflow one can spend several hours optimizing the tip, or choose to terminate quickly and make a new tip. The latter takes more effort, but continuing to try to optimize a poor tip is not likely to be successful (thus the value is significantly lower). Hence, enabling automated workflow planning and predictive optimization can significantly accelerate the experiments.

**IV. Are autonomous ML workflows practical?**

The implementation of machine learning driven automated experiments requires three components. The first and obvious one is the engineering control – meaning that ML agent should be able to execute the commands on the microscope. While a primary limiting factor for decades, the addition of Python APIs by major instrument manufacturers alleviates this problem. The second is the ML algorithms for data analytics and decision making. By now, the algorithms for active learning based on Bayesian Optimization and Reinforcement learning are well known, with multiple open code repositories and textbooks available. However, largely unexplored is the issue of the workflow planning – meaning definition of the reward or value functions.

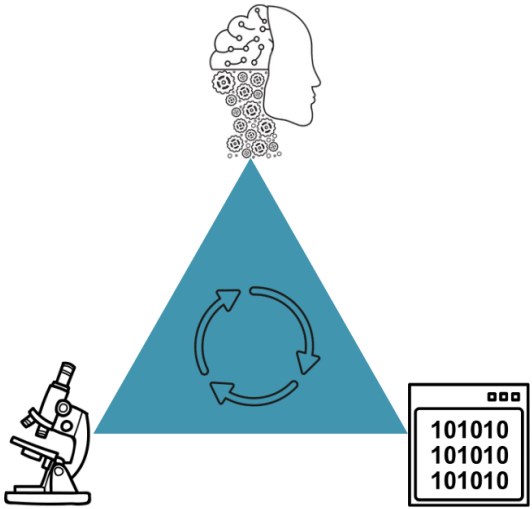

**Figure 4.** Three elements of the automated experiment. Engineering controls, ML algorithms, and workflow design.

Here, the key consideration becomes balancing the workflow flexibility with the data requirements. For example, the direct workflows relying on the identification of a priori known objects of interest requires robustness with respect to the changes in imaging parameters including resolution and sampling. This is a classic example of the out of distribution drift problem common in computer vision.

For workflows based on the concept of a reward function, the key consideration becomes the reward definition. Maximizing resolution is a well-defined reward that can be immediately obtained from the image. Comparatively, rewards reflecting materials discovery and optimization can be more complex. This is particularly the case for the sparse rewards becoming available upon completion of several experimental steps or at the end of experiment. In this case, a potential problem can be reward hacking, meaning the pursuance of the nominally correct reward that fails to align with the experiment objective. It is also important to note that real world actions are associated with the risks to the instrument – and hence an addition of "guardian angel" module to address and predict those can be required.

These constraints must be balanced with the limited experimental budgets. For example, RL workflows are remarkably data intensive. Bayesian Optimization methods are considerably data sparse, but the early observations suggest that the BO workflows can easily get trapped in the local minima. Hence, fully automated workflows are may not be optimal for automated experiments. This conclusion is not surprising – compare this to autonomous cars or use of ML in radiology, where the early optimistic prognoses have not yet been realized.

**V. hAE: From human running microscope to human in the loop**

We pose that workflow design for the microscopy in the near term will be a hybrid approach with the human in the loop. Here, the role of ML algorithm will be to perform fast low-level decisions, whereas the role of human operator will be to demonstrate the research objective, oversee the progression of the experiment and introduce the corrections for the ML operation, and assess experiment results. In other words, ML will cover the data generation and data complexity gap between the human perception and intrinsic instrument data generation rates.

We illustrate the concept of the hAE experiment using the example of a deep kernel learning (DKL) workflow recently introduced by our group,[3, 10] the same logic can be applied to other Bayesian Optimization workflows. The DKL workflow discovers the relationship between microstructural elements measured from imaging data and the certain functionalities measured from the spectroscopic data. Scalarizer functions are used to transform spectrum into the measure of functionalities. The operator can define the scalarizer functions based on what aspect of spectroscopic data would be of interest, e.g., it can be spectral intensity, peak width, peak position, or any other transform of the spectral data. The key aspect of scalarizer function is that it is immediately available in the experiment and is thus a myopic reward function.

The DKL workflow starts with the acquisition of the overview structural image. The image is sampled over the rectangular grid to produce a collection of image patches defining the local microstructure. This step identifies the collection of objects of interest and corresponding coordinates. With this input, spectroscopic measurement is performed at one or a few locations. The DKL is trained with the structural patches as input and the scalarizer functionality from spectrum as output, thus trained DKL makes predictions and uncertainties of functionality corresponding to each patch. This prediction and uncertainty can be combined into the classical acquisition functions (e.g., EI, UCB) to derive the next measurement location. With the next spectral measurement, DKL is retrained and the process is continued until the experimental budget is exhausted.

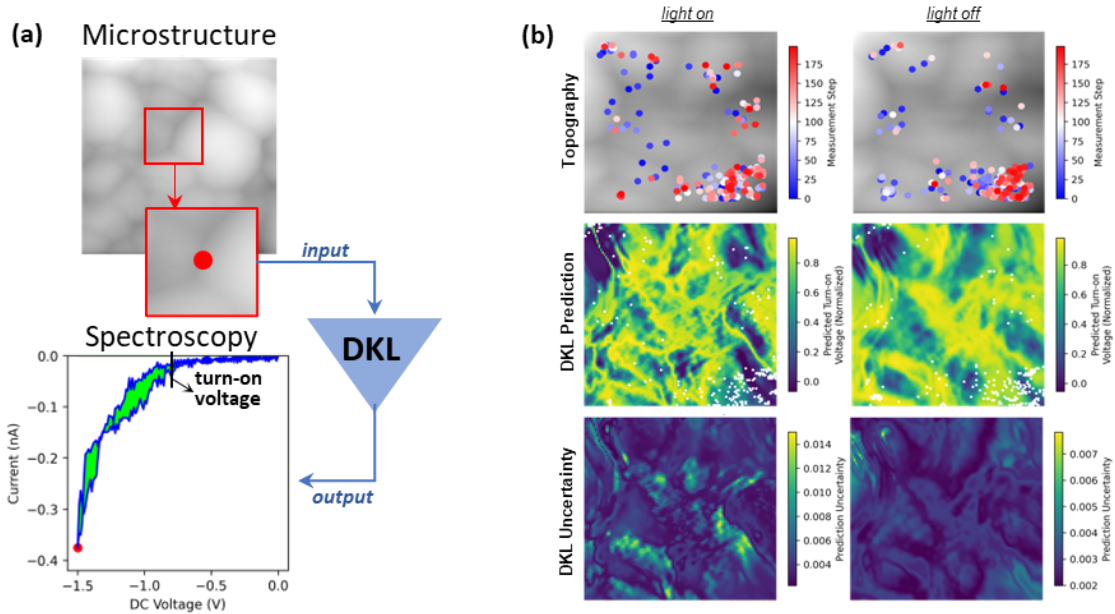

**Figure 5.** DKL exploration of structure-property relationship in hybrid perovskite, where the structure is topography showing grain and grain boundary, property is turn-on voltage extracted from I-V curve. The DKL exploration locations are mostly located around grain boundary junction points. Reproduced with permission from ref [11]

The DKL have shown remarkable efficiency in STEM-EELS, 4D STEM,[10] and multiple variants of the Scanning Probe Microscopy.[3, 12, 13] However, while in certain cases observed experimental progression was explainable, this was not always the case. The problem may be associated with the choice of scalarizer or policy, e.g., greedy policies are prone to trapping into local minima. Notably, the scalarizer and policies here are selected in advance and not changed by operators during experiments. We believe that human should be introduced in the loop of automated experiment workflows in microscopy, where the ML agent (e.g. DKL) issues the direct commands to the microscope at the intrinsic latencies of the microscope operation and the human operator monitors the progression of the automated experiment at human rates and tunes the behavior of the AI agent.

In the example of DKL microscopy and spectroscopy, the natural monitoring operations include the learning curve, i.e. the rate at which the DKL agent learns the relationship between the local structure and the scalarizer. Another monitoring function include real-space trajectory, and the trajectory in the feature space of the system. The latter can be defined as a latent space of the autoencoder trained on the full collection of the image patches. In the human in the loop experiment, the operator can issue commands affecting the behavior of the ML agent. This can induce the selection and tuning of the scalarizer function such as changing the selection of the spectral interval over which the EELS intensity is integrated, tuning the exploration-exploitation balance, either via the selection of the acquisition function or tuning its hyperparameter.

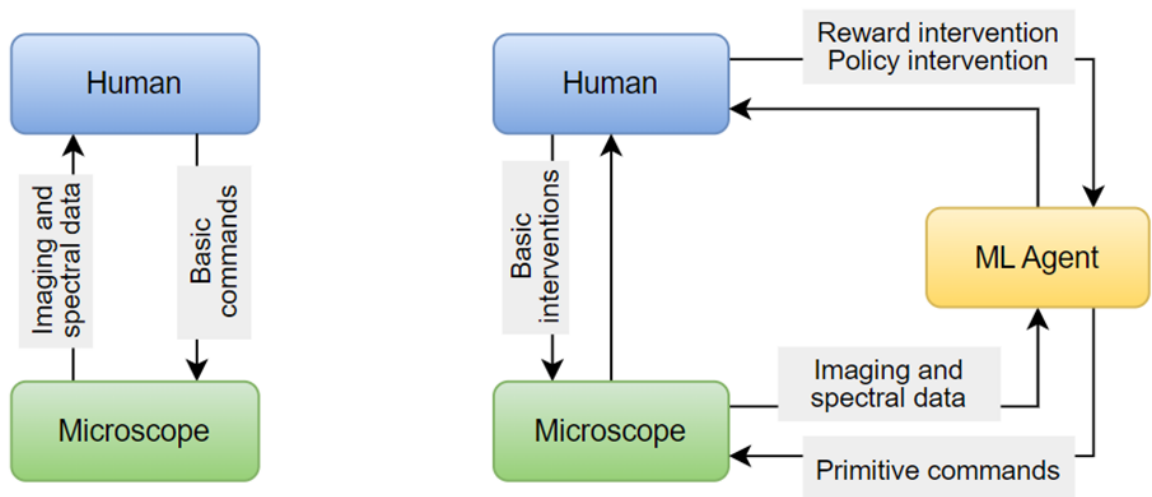

**Figure 6.** In the classical microscopy setting, the human directly assesses the data streamed from the microscope, makes decisions based on the analysis of this data, and issues commands to the microscope. In hAE approach, the ML agent collects the information form the microscope and issues commands to the microscope. The human operator observes the process and adjusts the policies of the ML agent, or performs direct interventions.

One example of the hAE approach is Bayesian Optimized Active Recommender System (BOARS).[2] In this architecture, a partial human intervention is augmented to the traditional BO approach during the early experiments when the parameter space is entirely unexplored. The human operator, instead of predefining a scalarizer or policy, provides "on the fly" visual assessment of the experimental results to design the human-identified scalarizer (spectra structure). In other words, in BOARS, the policy (target spectra structure) is not selected in advance and tuned in the loop of the experiments through monitoring and assessments. Once the cost of human assessment is exhausted, the BOARS switches to characterize fully autonomously guided locations to speed up the learning of the region of curiosity driven human identified spectral structures over the material structure space. While using effective A/B testing approach, BOARS increases efficiency in regard to alignment to the objective of the operators with hAE and provides better experimental steering. In some respects, this is similar to reinforcement learning methods that depend on human assessments rather than pre-determined rewards.[14]

---

[2] https://arxiv.org/abs/2304.02484

**VI. Opportunities**

Currently, most manufacturers have or plan to add APIs in microscopes, and the codebase from fields like RL, computer vision, etc. is available. Therefore, there is no (or very small) technical barrier to deploy all the power of modern ML on electron and probe microscopes, leading to opportunity for ML workflows in microscopy assuming the data pipelining is handled sufficiently. Given nowadays microscopes can execute commands and acquire data much faster than humans can comprehend data and make decisions, we look forward to, e.g., for STEM, 10-100x more efficient research for ML-driven microscopes, this can result in new opportunities for STEM, SPM, etc., and even fundamentally new experiment types, such as tracing and measuring defects and atomic fabrication.

The introduction of ML workflows can further stimulate formation of open data and code ecosystems. Here we note that the electron microscopy community has a very robust tradition of developing simulation codes - and most of these developments (outside of several patent algorithms) are open. However, all the tasks downstream from experiment - including data sharing, development of codes for data analytics, code base for automated experiment - tend to be done internally. We believe that the reason for that is that in most scientific workflows the electron microscopy images are the most downstream task - they are used to provide illustration of a certain phenomenon, prove a certain theory, or seed a theoretical model by initial structures. However, there are only very few cases (outside of CryoEM) of the downstream use of the data for materials design or learning transferable chemical and physical mechanisms.

The MSA 2023 has illustrated that the ML workflows are already becoming operationalized, e.g., for semiconductor industry, which has very clear experiment objectives that can be used to define reward, policy, and further workflows properly. However, ML workflows are still rarely explored in fundamental research fields, for which the development is often non-trivial and is outside of the reach of individual groups, and the experimental objectives can vary widely. The hAE workflows combine the power of ML and flexibility necessary for the human-lead exploratory research, and allow for seamless integration of human creativity and ML precision.

**Acknowledgements**

This effort was supported (SVK) as part of the center for 3D Ferroelectric Microelectronics (3DFeM), an Energy Frontier Research Center funded by the U.S. Department of Energy (DOE), Office of Science, Basic Energy Sciences under Award Number DE-SC0021118. A portion of this research was supported by the Center for Nanophase Materials Sciences (CNMS), which is a US Department of Energy, Office of Science User Facility at Oak Ridge National Laboratory. AB and MZ were supported US Department of Energy, Office of Science, Office of Basic Energy Sciences, MLExchange Project, award number 107514